\documentclass[conference]{IEEEtran}
\IEEEoverridecommandlockouts
\usepackage{cite}
\usepackage{amsmath,amssymb,amsfonts}
\usepackage{algorithmic}
\usepackage{graphicx}
\usepackage{textcomp}
\usepackage{xcolor}
\usepackage{array}
\usepackage{vcell}
\usepackage{multirow}
\usepackage{collcell}
\usepackage{hhline}
\usepackage[export]{adjustbox}
\usepackage[font=small]{caption}
\usepackage{subcaption}
\usepackage{graphicx}
\usepackage{tabularx}
\usepackage{booktabs}

\newcolumntype{C}[1]{>{\centering\arraybackslash}m{#1}}

\newcommand{\ru}{\rule{0mm}{3mm}}

\def\BibTeX{{\rm B\kern-.05em{\sc i\kern-.025em b}\kern-.08em
    T\kern-.1667em\lower.7ex\hbox{E}\kern-.125emX}}
\begin{document}

\title{Training-Free Deepfake Voice Recognition by Leveraging Large-Scale Pre-Trained Models}

\author{\IEEEauthorblockN{Alessandro Pianese, Davide Cozzolino, Giovanni Poggi, Luisa Verdoliva}
\IEEEauthorblockA{University of Naples Federico II\\
\{name.surname\}@unina.it}
}

\maketitle

\begin{abstract}
Generalization is a main issue for current audio deepfake detectors, which struggle to provide reliable results on out-of-distribution data.
Given the speed at which more and more accurate synthesis methods are developed, it is very important to design techniques that work well also on data they were not trained for.
In this paper we study the potential of large-scale pre-trained models for audio deepfake detection, with special focus on generalization ability.
To this end, the detection problem is reformulated in a speaker verification framework and fake audios are exposed by the mismatch between the voice sample under test and the voice of the claimed identity.
With this paradigm, no fake speech sample is necessary in training,
cutting off any link with the generation method at the root, and ensuring full generalization ability.
Features are extracted by general-purpose large pre-trained models, with no need for training or fine-tuning on specific fake detection or speaker verification datasets.
At detection time only a limited set of voice fragments of the identity under test is required.
Experiments on several datasets widespread in the community show that detectors based on pre-trained models achieve excellent performance and show strong generalization ability,
rivaling supervised methods on in-distribution data and largely overcoming them on out-of-distribution data.
\end{abstract}


\section{Introduction}
As deepfake editing tools become more and more accessible, manipulated content begins spreading widely on social networks.
In this context, voice deepfakes are playing a key role in the surge of misinformation because a spoofed recording can be just as effective at fooling people as a video deepfake is, but it is much easier to generate.
In a recent well-known episode, malicious actors used AI to impersonate a German chief executive’s voice on the phone, obtaining a fraudulent money transfer of 220,000 dollars\footnote{https://www.wsj.com/articles/fraudsters-use-ai-to-mimic-ceos-voice-in-unusual-cybercrime-case-11567157402}. Even more worrisome, voice deepfake can be used to spread disinformation in political campaigns, to defame people, arouse racial hatred, and a wide range of other illicit purposes.
This justifies the fast growing attention in recent years for voice deepfakes and their detection.
Actually, the interest on anti-spoofing methods is old with a large scientific literature that was mostly focused on Text to Speech (TTS) synthesis and Voice Conversion (VC) with classical methods.

Some papers rely on the extraction of acoustic features in the Fourier domain.
In fact, it has been recognized that synthesized speech exhibits unusual correlations between magnitude and phase spectra at the high frequencies \cite{albadawy2019detecting} and also that they exhibit an anomalous energy distribution among vocal formants \cite{cuccovillo2023audio}.
These model-based approaches provide not only a good performance but also high interpretability \cite{Salvi2023towards}.
Detectors based on handcrafted features, however, depend too much on the artifacts they are looking for.
To avoid this problem, in \cite{Tak2021end} a deep neural network classifier is developed that works directly on the raw speech signal. The first layer of the network is constrained to act as a filter bank, with filters of fixed shape and learned cut-in and cut-off frequencies.
Other solutions have been proposed to avoid over-fitting to specific artifacts, such as using features defined in different domains, {\it e.g.,} the temporal and spectral intervals \cite{Tak2021end2end-spectro, jung2022aasist, borrelli2021synthetic},
exploiting the limited ability of synthetic generators to perfectly reproduce the vocal tract \cite{blue2022you} or replicate certain phonemes, or using features related to the human articulation system \cite{Dhamyal2021using}.

The main problem with such approaches is that they cannot cope with the large variety of synthetic generators that are proposed day after day.
In fact, most of them use supervised learning and are not able to generalize to unseen attacks \cite{muller2022does}.
A possible solution is to follow a one-class approach where training is carried out only on the real class (bona fide speech) and data that depart from the learned model are classified as anomalies \cite{Zhang2021one-class}.
As fakes are not used during training, generalization would be automatically ensured.
It goes without saying that accurately modeling the actual voice class, with all its myriad variations, is simply out of reach.
However, following this line of thinking, it is possible to reframe the problem of  detection as a speaker verification task.
That is, instead of investigating whether or not a specific segment of speech was spoken by a human and not by a machine,
a simpler question can be asked: was it uttered by {\it this} specific human being?
Does this voice really belong to the person it is said to belong to?
This is obviously a different task, speaker verification, but in practice it implies voice deepfake detection.
In fact, if the claimed speaker is not recognized, there is good reason to suspect that the audio was manipulated and, at the very least, proceed with further and more thorough analyses.
On the other hand, recognizing the identity of the speaker is essential in order to give credence to the message conveyed by any speech. Who would give credit to messages coming from unknown people?
Not by chance, the first method designed to recognize deepfake videos with this approach aimed at ``protecting world leaders'' \cite{Agarwal2019} that is, very influential and well-known persons, which led to the so called person-of-interest (PoI)-based deepfake detectors \cite{cozzolino2021idreveal, bohavcek2022protecting}.
Note that these methods are different from \cite{Castan2022speaker, ding2023samo}, where a large dataset of real/synthetic audios is needed for training.
It is also interesting to highlight an exchange of roles. Here speaker verification is used as a means to detect forgeries, while usually, in biometric access systems, forgery detection is used as a preliminary screening before proceeding to speaker recognition \cite{Shim2022baseline}.
Of course, our approach also relies on a few hypotheses.
We assume that we know the identity of the speaker for the audio under analysis and, more importantly, that we have a reference set of pristine audio tracks for that identity.
This scenario is similar to some previous works \cite{pianese2022deepfake, cozzolino2023audiovisual} where, however, the model must be explicitly trained for identity recognition.

\begin{figure}[t!]
    \centering
    \includegraphics[width=.85\linewidth,clip, trim=45 110 45 0]{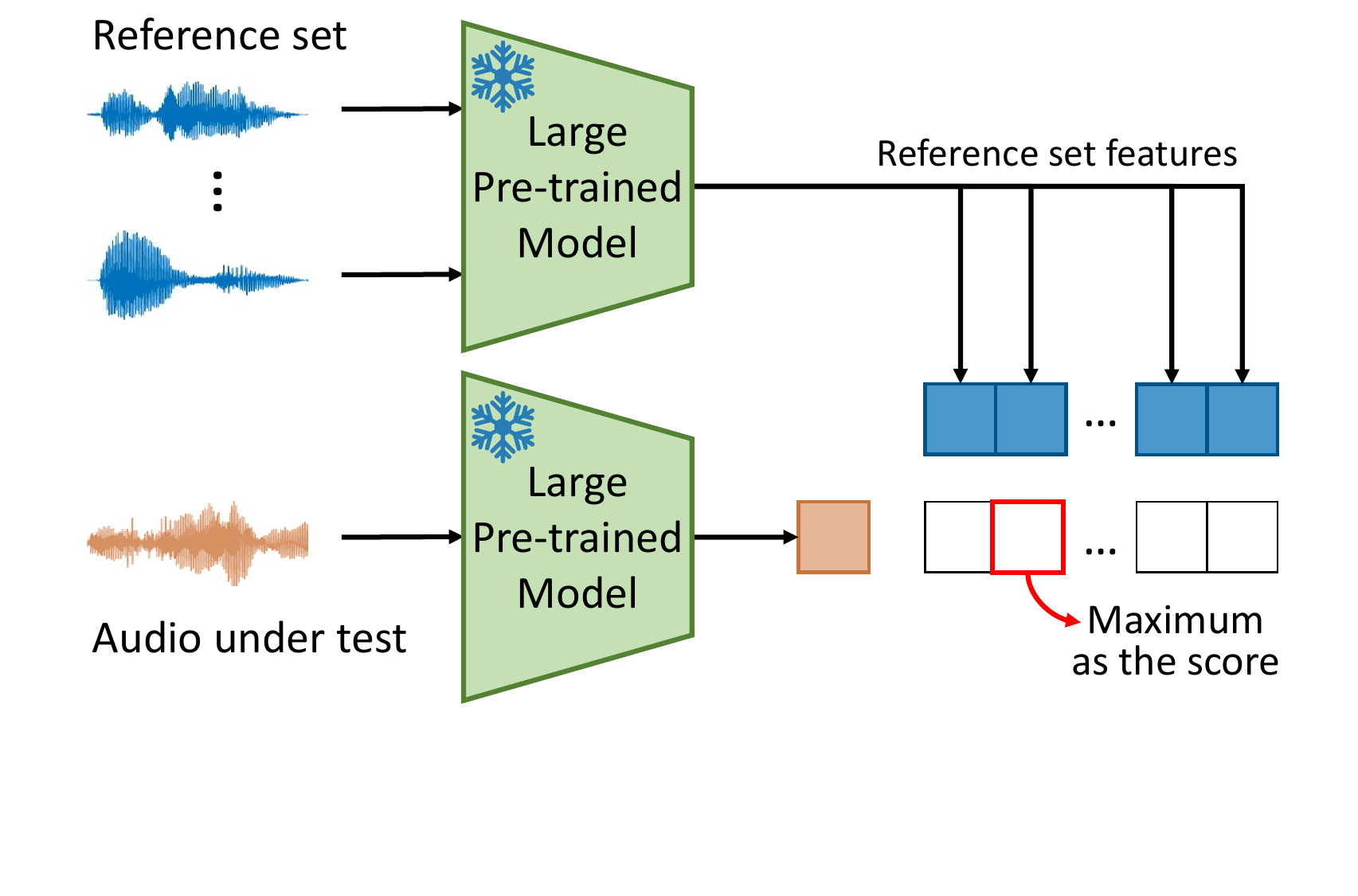}
    \caption{Making decisions based on large pre-trained models. Audio signals are fed to the model to extract the corresponding embeddings. The decision statistic is computed in the latent space, as the maximum similarity score between the audio under test and the audios of the reference set.}
    \label{fig:block_scheme}
\end{figure}

In this paper we move an important step forward,
leveraging the power of large pre-trained models to develop a solution that is fully training-free.
A few recent papers follow a similar path \cite{barrington2023single, Oneata2023towards} but,
while relying on pre-trained features, they still need to train a classifier on a suitable dataset of real and synthetic data.
In our method, instead, any dependency on fake samples is removed, ensuring the highest generalization.

\section{Proposed Framework}

We consider a scenario in which
{\it  i)} the identity of the speaker is known and
{\it ii)} a reference set, $\mathcal{R}$, of real audios of this same identity is available.
Both hypotheses are certainly reasonable and easily met.
It is difficult to imagine situations in which anonymous audio would be relevant to the point of having to verify its authenticity.
Furthermore, in the age of social networks, many audio samples are usually available for any individual, and in any case, a person who is the subject of a possible deepfake may be more than willing to generate new samples to prove or disprove the attack.
For example, a person accused of uttering sexist or racist phrases off-air can prove the audio snippet is a deepfake by releasing a large collection of new phrases for reference.
The comparison between the audio under test and those coming from the reference set is carried out in the embedding space.
The verification/detection procedure is outlined in Fig.\ref{fig:block_scheme}.
Both the audio under test (orange) and the reference samples (blue)
are provided as input to a pre-trained model to obtain the corresponding features in the embedding space,
$x$ and $\left\{r_i, i=1,\ldots,|\mathcal{R}|\right\}$, respectively.
Then, similarity scores $s_i=sim(x,r_i)$ are computed, using cosine-based similarity.
Finally, the maximum similarity score $s=\max \left\{ s_i \right\}_{i\in \mathcal{R}}$ is taken as the decision statistic
and compared to an appropriate threshold to label the audio as real (the claimed identity is confirmed) or fake.

The maximum similarity criterion takes into account the intrinsic variability of speech samples generated by a single speaker, which can lead to pretty widespread features in the embedding space.
To be validated as real, the test audio only needs to be very similar to one of the certified reference samples, while the requirement to be similar to all of them could be too stringent and ultimately misleading, an intuition also confirmed by the experiments carried out in \cite{cozzolino2021idreveal,cozzolino2023audiovisual}.
For this reason, it is very important to have an adequate number of different speech samples in the reference, so as to cover the entire region of the embedding space in which features of the same speaker can occur.
On the other hand, as the number of reference samples grows, one of them might happen by chance to be too close to the deepfake under test, and the latter could be classified as real.
The dependence of performance on the reference set size and on the decision threshold will be explored experimentally in subsection 3.3.

In this work, we choose not to design or re-train a model to extract dedicated features for the problem at hand,
but rather to use generic features extracted from large pre-trained audio models.
In Fig.\ref{fig:block_scheme} this choice is represented by the ice symbol on the feature extractor, which means it is taken off-the-shelf, with frozen weights, and no training is required at any stage of the process.
This choice is motivated by the impressive performance exhibited by large pre-trained models in many downstream applications, including some in image forensics \cite{ojha2023towards,cozzolino2023raising},
and by the clear advantage of having a training-free solution, thus eliminating any dependence on training sets and their inevitable polarizations.

\begin{table}[t!]
    \centering
    \resizebox{\columnwidth}{!}{
    \begin{tabular}{lccC{0.1\textwidth}c}
    \toprule
    Model & Type & \# Params & Speech (h) & Task \\  \midrule
    Wav2Vec2-xlsr & CNN + Transf. &   2B & 436K & SR\\
    AudioCLIP     & CNN + Transf. & 134M &   5K & LRL\\
    LaionCLAP     & Transformer   & 158M &   9K & LRL\\
    BEATs         & Transformer   &  90M &   5K & LRL \\
    \bottomrule
    \end{tabular}
    }
    \vspace{2mm}
    \caption{Pre-trained models used in our analysis with information on the type of architecture, the number of parameters, the hours of speech in training and the training task:  Speech Recognition (SR) and Latent Representation Learning (LRL).}
    \label{tab:models}
\end{table}

\renewcommand{\t}[1]{\textbf{#1}}
\renewcommand{\v}[1]{#1}
\setlength{\tabcolsep}{4.5pt}
\begin{table*}[t!]
    \centering
    \begin{tabular}{lrrrrrrrrrrrrr}
    \toprule
\ru                                           &  \multicolumn{3}{c}{ASVSpoof2019} & \multicolumn{3}{c}{ASVSpoof2021} & \multicolumn{3}{c}{InTheWild} & \multicolumn{3}{c}{Average} & \multicolumn{1}{c}{$\sigma$} \\ \cmidrule(lr){2-4} \cmidrule(lr){5-7} \cmidrule(lr){8-10} \cmidrule(lr){11-13} \cmidrule(lr){14-14}
\ru Method                                    &      EER &    t-DCF &      AUC &    EER   &    t-DCF &      AUC &    EER   &    t-DCF &      AUC &      EER &    t-DCF &      AUC &      AUC \\ \midrule
\ru RawNet2 \cite{Tak2021end}                 &    18.21 &    0.595 &    88.49 &    22.33 &    0.700 &    82.79 &    49.53 &    0.999 &    50.68 &    30.02 &    0.765 &    73.99 &    16.64 \\
\ru RawGATstMul \cite{Tak2021end2end-spectro} &     1.07 &    0.031 &    99.90 &    12.52 &    0.211 &    92.07 &    52.53 &    0.994 &    47.36 &    22.04 &    0.412 &    79.78 &    23.14 \\                  
\ru AASIST \cite{jung2022aasist}              & \t{ 0.83}&    0.027 & \t{99.92}&    11.20 &    0.231 &    96.16 &    43.01 &    0.901 &    60.12 &    18.35 &    0.386 &    85.40 &    17.94 \\                  
\ru TitaNet+LR \cite{barrington2023single}    &    16.04 &    0.499 &    83.96 &    21.87 &    0.593 &    78.13 &    46.74 &    0.945 &    53.26 &    28.22 &    0.679 &    71.78 &    13.31 \\                  
\ru SAMO \cite{ding2023samo}                  &     0.88 & \t{0.026}&    99.71 &    12.09 &    0.212 &    94.74 &    46.94 &    0.952 &    55.17 &    19.97 &    0.397 &    83.21 &    19.92 \\ \hline           
\ru WavLM \cite{chen2022wavlm}                &    15.12 &    0.493 &    92.58 & \t{ 0.00}& \t{0.000}& \t{99.99}&    23.52 &    0.597 &    84.30 &    12.88 &    0.363 &    92.30 &     6.40 \\                  
\ru ECAPA-TDNN \cite{desplanques2020ecapa}    &    20.04 &    0.682 &    86.52 &     4.74 &    0.159 &    99.09 &    27.37 &    0.848 &    76.60 &    17.39 &    0.562 &    87.40 &     9.20 \\                  
\ru TitaNet \cite{koluguri2022titanet}        &    14.17 &    0.456 &    93.61 &     0.02 & \t{0.000}& \t{99.99}&    24.10 &    0.645 &    83.77 &    12.76 &    0.367 &    92.46 &     6.67 \\                  
\ru POI-Audio \cite{pianese2022deepfake}      &    11.50 &    0.358 &    95.34 &     0.37 &    0.006 &    99.94 &    14.73 &    0.379 &    92.06 &     8.87 &    0.248 &    95.78 &     3.23 \\ \hline           
\ru Wav2Vec2-xlsr-2B                          &     8.65 &    0.244 &    96.68 &     5.15 &    0.101 &    97.98 &    19.19 &    0.550 &    86.95 &    11.00 &    0.298 &    93.87 &     4.92 \\                  
\ru AudioCLIP                                 &    32.09 &    0.845 &    75.85 &     2.98 &    0.076 &    99.52 &    38.52 &    0.998 &    64.93 &    24.53 &    0.640 &    80.10 &    14.43 \\                  
\ru LaionCLAP                                 &    16.61 &    0.488 &    91.14 &     3.74 &    0.109 &    99.46 &    11.76 &    0.348 &    95.16 &    10.77 &    0.315 &    95.25 &     3.39 \\                  
\ru BEATs                                     &     2.73 &    0.079 &    99.61 &     0.86 &    0.027 &    99.97 & \t{ 1.62}& \t{0.035}& \t{99.73}& \t{ 1.74}& \t{0.047}& \t{99.77}&     0.14 \\                  
\hline                                              
    \end{tabular}
    \vspace{2mm}
    \caption{Comparison with SOTA in terms of EER, t-DCF, AUC.
    The first section of the table reports scores for supervised models trained on ASVSpoof2019. 
    The second section is dedicated to speaker verification methods. 
    The bottom section gives the performance of the proposed approach with different large pre-trained models.
    The last four columns report the average values for all metrics and, only for the AUC, also the standard deviation ($\sigma$).
    For each column, the best result is in bold.}
    \label{tab:detection_comparison}
\end{table*}

Here we consider and compare several popular pre-trained models. They are described below, while a summary of their characteristics is reported in Table \ref{tab:models}.
\textbf{Wav2Vec2-xlsr} \cite{babu2021xls}
is a large-scale pre-trained model for speech representation learning.
It is based on  wav2vec 2.0 \cite{baevski2020wav2vec} and trained on nearly half a million hours of publicly available speech audio, an order of magnitude more data than previous models.
Its main peculiarity is the focus on cross-lingual training, spanning 128 different languages from different world regions, which turns out to be important for some applications.
\textbf{AudioCLIP} \cite{guzhov2022audioclip}
extends the CLIP model \cite{radford2021learning}, originally proposed for text-image representation alignment, by adding ESResNeXt \cite{guzhov2021esresne} as audio head.
Through this addition, the model is trained to learn aligned text-image, text-audio, and audio-image representation.
The trained model allows the base architecture to operate also with the audio modality
and enables cross-querying using images, text, and audio.
\textbf{LaionCLAP} \cite{laionclap2023}
is a large model trained on the LAION-Audio-630 dataset containing 633526 audio-text pairs collected from different sources.
It implements a CLIP-like training strategy with image data replaced by audio data:
audio and text data are fed to the network and features of correlated data are brought together while those of unrelated data are spaced apart.
Audios of different lengths can be dealt with seamlessly.
Finally, \textbf{BEATs} \cite{chen2023beats}
proposes an innovative iterative pre-training framework for audio representation learning.
The proposed architecture includes both a tokenizer and an audio self-supervised learning model
which are trained jointly through a series of iterations, with each model contributing to the other's learning process.
In the end, semantically rich tokens are extracted along with insightful representations of the audio.

\section{Experimental results}

For our experiments we consider the datasets that were proposed for the 2019 and 2021 editions of the Automatic Verification Spoofing and Countermeasure challenge, namely
\textbf{ASVSpoof2019} \cite{wang2020asvspoof} and \textbf{ASVSpoof2021} \cite{yamagishi2021asvspoof}.
The former contains 7,355 real and 63,882 spoofed audios generated using 13 different speech synthesis methods and voice conversion models, while the latter contains 22,617 real tracks and 589,212 fake ones.
We use only a subset of this dataset for which the speaker identity is known, for a total of 159,696 audio files.
To simulate a real use case, the DF audio tracks are compressed and decompressed using general purpose audio compression tools.

We also use the \textbf{InTheWild} dataset \cite{muller2022does}, comprising 11,816 fake and 19,963 pristine audios and aiming at representing a real-world scenario.
The dataset includes 54 identities, with real audios selected to closely match the corresponding fakes (e.g. similar speaker emotion, background noise, duration etc.).
Since it is collected in the wild, this dataset provides no information on the techniques used for generation.

To measure performance we consider the metrics usually adopted in the specific literature:
the equal error rate (EER) which is the point on the receiver operating characteristic curve (ROC) where false acceptance rate and false rejection rate are equal;
the tandem detection cost function (t-DCF) which is a weighted sum of missing detection rate, false alarm rate, and  prior class probabilities;
and the area under the ROC curve (AUC).

\subsection{Comparison with the state-of-the-art}
We compare our method with various state-of-the-art approaches from the current literature.
We consider the following supervised methods: RawNet2 \cite{Tak2021end}, RawGATstMul \cite{Tak2021end2end-spectro}, AASIST \cite{jung2022aasist}, Titanet + LR \cite{barrington2023single}, SAMO \cite{ding2023samo}, all trained on the fake/real ASVSpoof2019 dataset.
Then, we consider unsupervised approaches that work under the same hypothesis as ours, that is, they only assume to know a reference set of pristine audios
(from 3 to 711, in this experiment, depending on the speaker, with an average value of 240).
This set includes some speaker verification methods: WavLM \cite{chen2022wavlm}, ECAPA-TDNN \cite{desplanques2020ecapa} and TitaNet \cite{koluguri2022titanet}, and also a speech deepfake detection method, POI-Audio \cite{pianese2022deepfake} pre-trained on real audios of more than 5,000 identities.

All results are reported in Table \ref{tab:detection_comparison}.
As expected, supervised approaches, trained on the training partition of ASVSpoof2019, perform very well in the perfectly aligned scenario, that is, on ASVSpoof2019, 
but the performance worsen significantly as soon as new datasets are considered, to collapse altogether on the more challenging InTheWild dataset, with coin-tossing results except for AASIST, which shows an AUC just above $60\%$.
Methods based on the prior knowledge of the individual under test (second section) are much more stable, with top performance on ASVSpoof2021 dataset and competitive results on the others, including InTheWild.
The average AUC is always around $90\%$.
Very similar results are observed also for methods of the third section, based on large-scale pre-trained models. 
This confirms the potential of the identity-based approach for deepfake detection, especially in terms of generalization, even in the absence of any specific training for the speaker verification or identity recognition tasks.
To further underline this point, in the last column of the table we also report the standard deviation of the AUC, by which a clear gap between identity-based and supervised methods appears, with the only notable exception of AudioCLIP.
The case of BEATs (last row) deserves some more words.
Average results speak by themselves: they are the best under all metrics, with  $1.74\%$ EER (second best $8.87\%$), $0.047$  t-DCF (second best $0.248$), and $99.77$ AUC (second best $95.78$).
In fact, they are not only very good, but also extremely stable across all datasets, as also obvious by the very small standard deviation of the AUC shown in the last column.
The quality of its representation, based on a semantic-rich acoustic tokenizer, appears to be much superior to that of other pre-trained models, opening the way to excellent results in this application.

\subsection{Representation quality of large models}

\begin{figure}[t!]
    \centering
    \begin{subfigure}[t]{0.49\linewidth}
        \includegraphics[width=\linewidth]{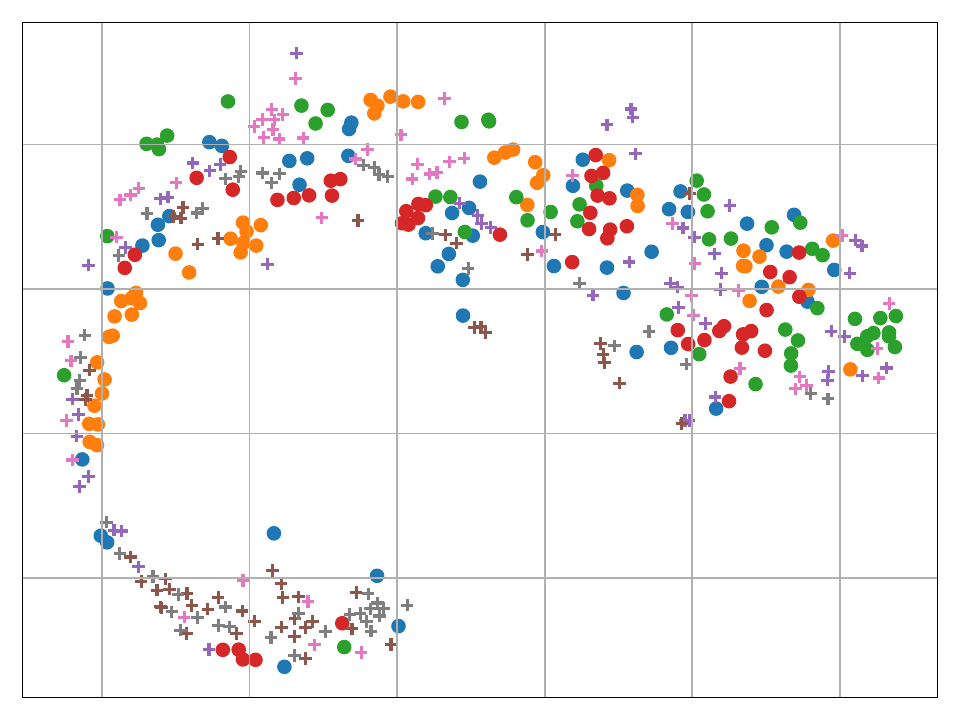}
        \vspace{-5mm} \caption{Wav2Vec2-xlsr}
        \label{fig:tsne_wav2vec2}
    \end{subfigure}
    \hfill
   \begin{subfigure}[t]{0.49\linewidth}
        \includegraphics[width=\linewidth]{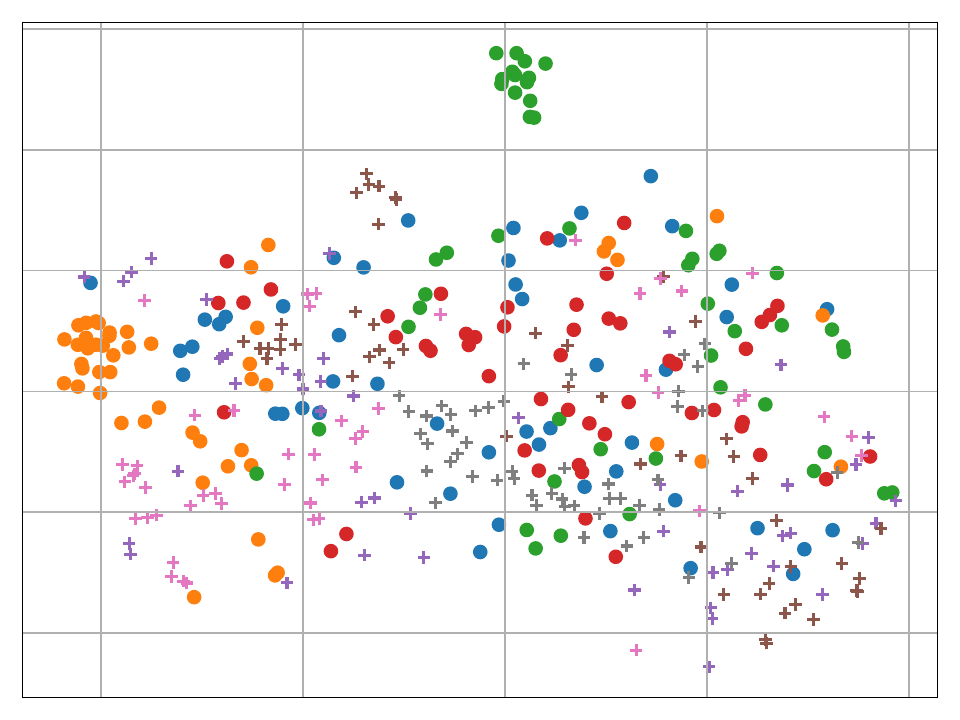}
        \vspace{-5mm} \caption{AudioCLIP}
        \label{fig:tsne_audioclip}
   \end{subfigure}

    \begin{subfigure}[t]{0.49\linewidth}
        \includegraphics[width=\linewidth]{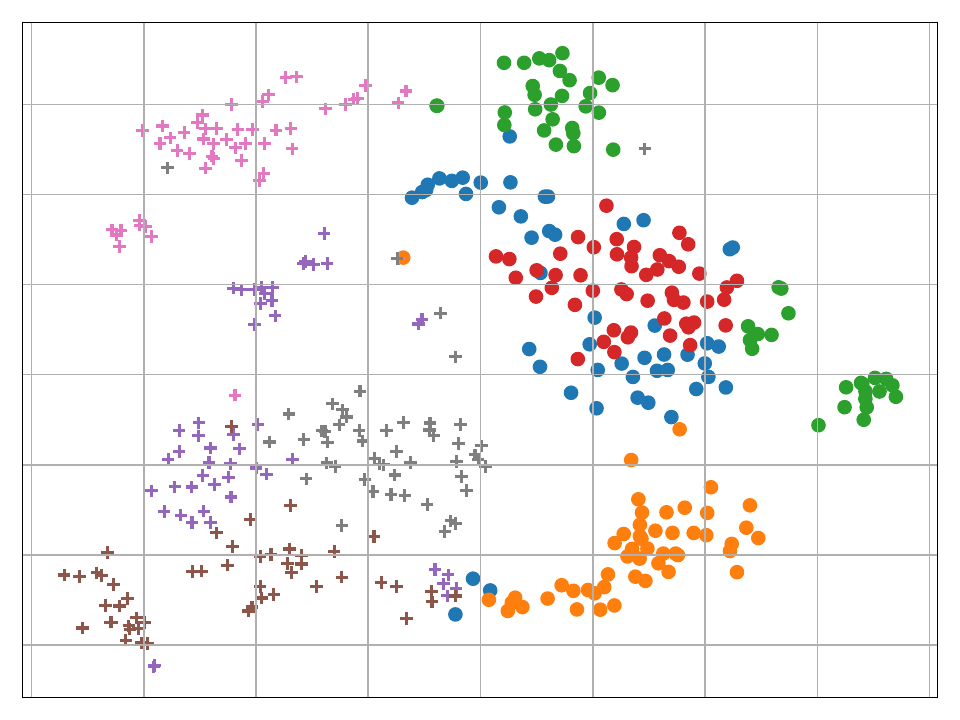}
         \vspace{-5mm} \caption{LaionCLAP}
        \label{fig:tsne_laionclap}
    \end{subfigure}
    \hfill
    \begin{subfigure}[t]{0.49\linewidth}
        \includegraphics[width=\linewidth]{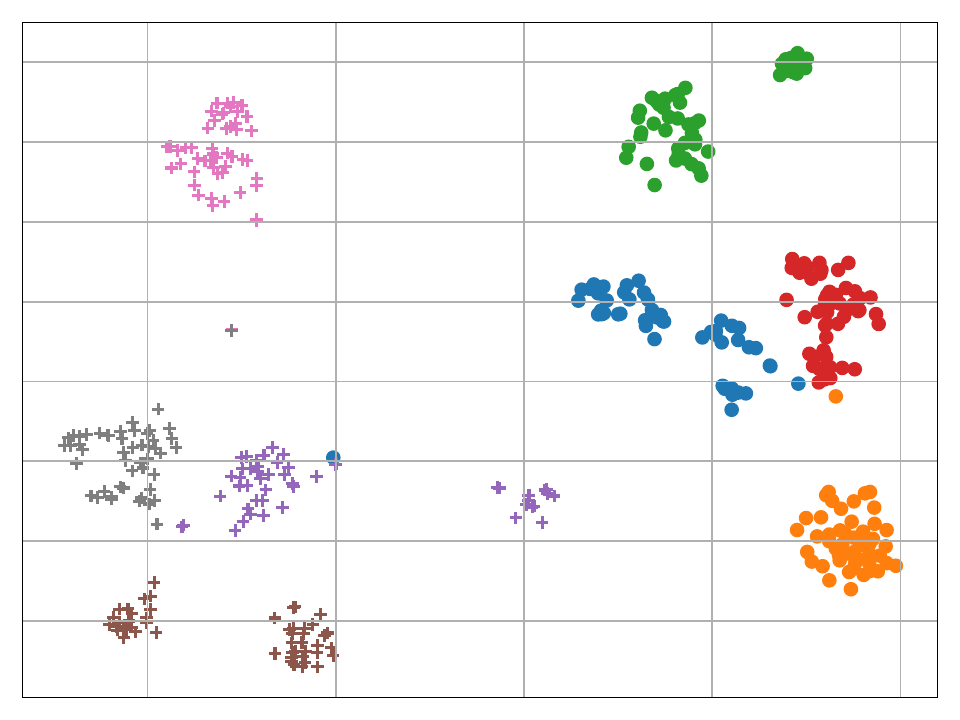}
        \vspace{-5mm} \caption{BEATs}
        \label{fig:tsne_beats}
    \end{subfigure}

   \adjincludegraphics[width=0.9\linewidth,trim={{0.1\width} {.33\width} {.1\width} {0.31\width}},clip]{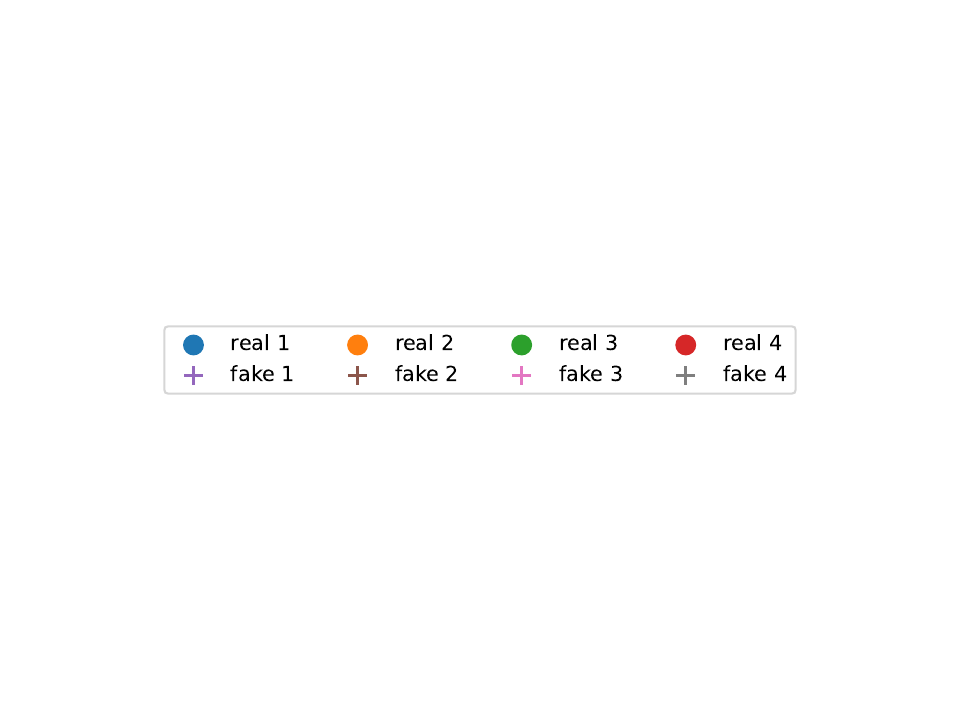}
   \caption{T-SNE representation of real and fake feature embeddings of four identities taken from the InTheWild dataset.}
   \label{fig:tsnePlot}
\end{figure}

The experimental results discussed before confirm the prevailing importance of a good learned representation, based on a large suitably pre-trained model, over specific adjustments of the detection method.
To better investigate this point, and explain the impact of representation quality on performance, in Figure \ref{fig:tsnePlot} we show a scatter plot of the 2D projections of 50 embedded vectors for each class.
Such projections are obtained by means of the t-SNE dimensionality reduction technique \cite{Van2008visualizing}.
Vectors are extracted from real and synthetic voices of four identities from the InTheWild dataset.
For Wav2Vec2 and especially for AudioCLIP it is difficult to recognize clusters associated with specific speakers, nor is it possible to separate real samples from fake ones.
Although these are only 2D projections, they indicate a poor representation quality, which explains some disappointing results.
On the contrary, with LaionCLAP definite clusters begin to emerge, which then become extremely compact and well-separated with BEATs.
They allow to easily distinguish each speaker from all the others as well as real audio from fake audio, explaining the excellent performance observed in Tab.\ref{tab:detection_comparison}.

It seems clear that BEATs ensures a largely improved quality of representation so we dedicate a few more words to describe it and possibly learn useful lessons.
BEATs moves from the observation that successful image and video large models are trained to predict semantic-rich discrete labels, such as ``horse'', ``bicycle'', etc.
On the contrary, state-of-the-art audio large models are trained to minimize a reconstruction loss, like the mean square error, that focuses on redundant and perceptually useless details rather than high-level semantic aspects.
In other words, a measure of distortion that is only weakly correlated with the actual distortion perceived by the end user.
This happens because it is difficult to associate semantic labels to audio fragments.
The key step taken in BEATs is to train a semantic tokenizer {\em jointly} with the large audio model in an iterative modality.
A random acoustic tokenizer is used to train the model, which then distills semantic knowledge useful for training a better tokenizer, and so on in a virtuous circle.
Experimental results, including those presented here, demonstrate that this solution is very effective and that BEATs is able to provide audio representations with high semantic value.

\subsection{Impact of the reference set size}

\begin{figure}[t!]
    \centering
    \includegraphics[width=0.99\linewidth,trim=20 16 5 100, clip]{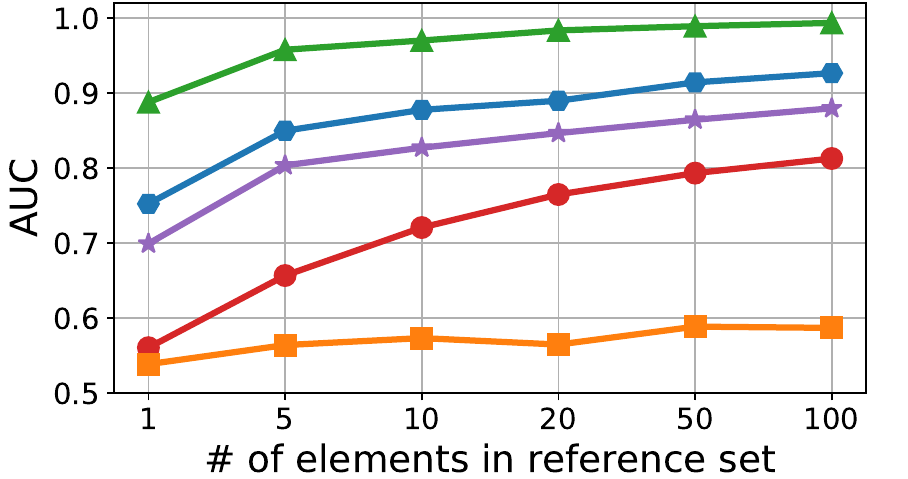} \\
    \includegraphics[width=0.99\linewidth,trim=0 255 0 30, clip]{img/reference_detection_lpm.pdf}
    \caption{AUC for the InTheWild dataset as a function of the number of audios in the reference set.}
    \label{fig:reference_detection}
\end{figure}

Finally, we study the impact of the reference set size on performance for the large pre-trained models described above and for the PoI-Audio method.
In Figure \ref{fig:reference_detection}, with reference to the InTheWild dataset, we report the AUC as a function of the number of speech segments available in the reference set for each identity under test. In all cases, the AUC improves steadily with increasing number of audios, but the ranking of methods remains the same over the whole range.
With BEATs, an excellent performance ($>0.9$) is achieved already with five audios, while LaionCLAP, the second best method, reaches a comparable level only with 100 samples.
Only AudioCLIP confirms to struggle, no matter how many reference audios are available, but this is easily explained by the poor representation quality observed in Fig.\ref{fig:tsnePlot}.

\begin{figure}[t!]
    \centering
    \includegraphics[width=0.32\linewidth,trim=0 0 0 0,page=1]{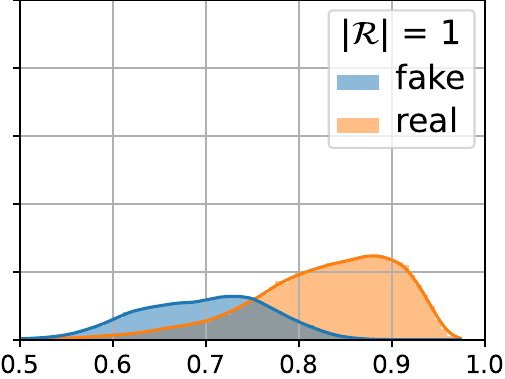}
    \includegraphics[width=0.32\linewidth,trim=0 0 0 0,page=2]{img/hist.pdf}
    \includegraphics[width=0.32\linewidth,trim=0 0 0 0,page=3]{img/hist.pdf} \\
    \includegraphics[width=0.32\linewidth,trim=0 0 0 0,page=4]{img/hist.pdf}
    \includegraphics[width=0.32\linewidth,trim=0 0 0 0,page=5]{img/hist.pdf}
    \includegraphics[width=0.32\linewidth,trim=0 0 0 0,page=6]{img/hist.pdf}
    \caption{Histograms of maximum similarity scores for BEATs on InTheWild dataset.}
    \label{fig:hist_th}
\end{figure}

\begin{figure}[t!]
    \centering
    \includegraphics[width=0.99\linewidth,trim=0 0 0 0]{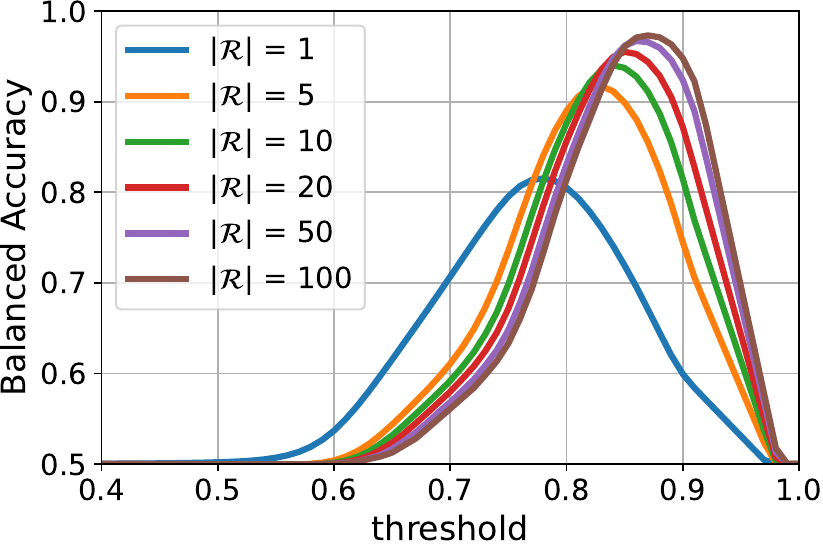}
    \caption{Accuracy of BEATs on InTheWild dataset as a function of threshold.}
    \label{fig:BEATs_thresholds}
\end{figure}

In Figure \ref{fig:hist_th}, only for the most interesting case of BEATs, we show the histograms of the maximum similarity scores for real and fake test samples (leave-one-out method) for various sizes of the reference set.
Starting from $|\mathcal{R}|=5$ the real and fake histograms are well separated.
As $|\mathcal{R}|$ grows,
The former becomes more and more concentrated at large values, as finding a very similar reference becomes progressively easier.
The latter, on the contrary, remains quite stable, which dispels the fear of accidentally finding very similar samples even when the test audio is fake, just because there are so many of them.
Of course, the optimal decision threshold changes as the reference set size increases, but only marginally so, as shown in Figure \ref{fig:BEATs_thresholds}, 
which allows to set the threshold in advance to a reasonable value, say, 0.85, that works well in all cases (again, except the case of a single reference).

\section{Conclusion}
Synthetic speech detection is becoming more and more relevant nowadays.
Conventional detectors, based on supervised learning, provide an excellent performance on data they were trained for, but suffer dramatic performance losses in out-of-distribution real-world scenarios, where the model used to generate fake audios is not known.
To ensure good generalization, we propose to use a one-class strategy, based on the only hypotheses to know the  speaker identity and  have a few reference audios at test time.
In this context, we analyze a training-free solution based on the use of large scale models pre-trained on audio-related tasks.
Results are extremely promising, competitive with supervised methods in aligned settings, and much superior to them in real-world scenarios. All this without any training on synthetic or field-specific data.
Among the other interesting results, our experimental analysis demonstrates the excellent performance of BEATs, a large audio model pre-trained to exploit semantic information, which suggests equally good results for other semantic tasks.

\section*{Acknowledgment}

We gratefully acknowledge the support of this research by a TUM-IAS Hans Fischer Senior Fellowship and the PNRR ICSC National Research Centre for High Performance Computing, Big Data and Quantum Computing (CN00000013), under the NRRP MUR program funded by the NextGenerationEU. This material is also based on research sponsored by the Defense Advanced Research Projects Agency (DARPA) and the Air Force Research Laboratory (AFRL) under agreement number FA8750-20-2-1004. The U.S. Government is authorized to reproduce and distribute reprints for Governmental purposes notwithstanding any copyright notation thereon. The views and conclusions contained herein are those of the authors and should not be interpreted as necessarily representing the official policies or endorsements, either expressed or implied, of DARPA or the U.S. Government. In addition, this work has received funding by the European Union under the Horizon Europe vera.ai project, Grant Agreement number 101070093.

\bibliographystyle{plain}
\bibliography{arXiv/bibliography}

\begin{thebibliography}{10}

\bibitem{Agarwal2019}
Shruti Agarwal, Hany Farid, Yuming Gu, Mingming He, Koki Nagano, and Hao Li.
\newblock {Protecting World Leaders Against Deep Fakes}.
\newblock In {\em IEEE Conference on Computer Vision and Pattern Recognition (CVPR) Workshops}, pages 38--45, 2019.

\bibitem{albadawy2019detecting}
Ehab~A AlBadawy, Siwei Lyu, and Hany Farid.
\newblock {Detecting AI-Synthesized Speech Using Bispectral Analysis}.
\newblock In {\em IEEE Conference on Computer Vision and Pattern Recognition (CVPR) Workshops}, pages 104--109, 2019.

\bibitem{babu2021xls}
Arun Babu, Changhan Wang, Andros Tjandra, Kushal Lakhotia, Qiantong Xu, Naman Goyal, Kritika Singh, Patrick von Platen, Yatharth Saraf, Juan Pino, et~al.
\newblock {XLS-R: Self-supervised Cross-lingual Speech Representation Learning at Scale}.
\newblock In {\em Annual Conference of the International Speech Communication Association (INTERSPEECH)}, pages 2278–--2282, 2022.

\bibitem{baevski2020wav2vec}
Alexei Baevski, Yuhao Zhou, Abdelrahman Mohamed, and Michael Auli.
\newblock {wav2vec 2.0: A Framework for Self-Supervised Learning of Speech Representations}.
\newblock In {\em Advances in Neural Information Processing Systems (NeurIPS)}, volume~33, pages 12449--12460, 2020.

\bibitem{barrington2023single}
Sarah Barrington, Romit Barua, Gautham Koorma, and Hany Farid.
\newblock Single and multi-speaker cloned voice detection: From perceptual to learned features.
\newblock In {\em IEEE international workshop on information forensics and security (WIFS)}, pages 1--6, 2023.

\bibitem{blue2022you}
Logan Blue, Kevin Warren, Hadi Abdullah, Cassidy Gibson, Luis Vargas, Jessica O'Dell, Kevin Butler, and Patrick Traynor.
\newblock {Who Are You (I Really Wanna Know)? Detecting Audio $\{$DeepFakes$\}$ Through Vocal Tract Reconstruction}.
\newblock In {\em 31st USENIX Security Symposium (USENIX Security)}, pages 2691--2708, 2022.

\bibitem{bohavcek2022protecting}
Maty{\'a}{\v{s}} Boh{\'a}{\v{c}}ek and Hany Farid.
\newblock Protecting world leaders against deep fakes using facial, gestural, and vocal mannerisms.
\newblock {\em Proceedings of the National Academy of Sciences}, 119(48):e2216035119, 2022.

\bibitem{borrelli2021synthetic}
Clara Borrelli, Paolo Bestagini, Fabio Antonacci, Augusto Sarti, and Stefano Tubaro.
\newblock Synthetic speech detection through short-term and long-term prediction traces.
\newblock {\em EURASIP Journal on Information Security}, 2021(1):1--14, 2021.

\bibitem{Castan2022speaker}
Diego Castan, Md~Hafizur Rahman, Sarah Bakst, Chris Cobo-Kroenke, Mitchell McLaren, Martin Graciarena, and Aaron Lawson.
\newblock {Speaker-targeted Synthetic Speech Detection}.
\newblock Technical report, Sandia National Lab.(SNL-NM), Albuquerque, NM (United States), 2022.

\bibitem{chen2022wavlm}
Sanyuan Chen, Chengyi Wang, Zhengyang Chen, Yu~Wu, Shujie Liu, Zhuo Chen, Jinyu Li, Naoyuki Kanda, Takuya Yoshioka, Xiong Xiao, et~al.
\newblock {WavLM: Large-Scale Self-Supervised Pre-Training for Full Stack Speech Processing}.
\newblock {\em IEEE Journal of Selected Topics in Signal Processing}, 16(6):1505--1518, 2022.

\bibitem{chen2023beats}
Sanyuan Chen, Yu~Wu, Chengyi Wang, Shujie Liu, Daniel Tompkins, Zhuo Chen, Wanxiang Che, Xiangzhan Yu, and Furu Wei.
\newblock {BEATs: Audio Pre-Training with Acoustic Tokenizers}.
\newblock In {\em International Conference on Machine Learning (ICML)}, pages 5178--5193, 2023.

\bibitem{cozzolino2023audiovisual}
Davide Cozzolino, Alessandro Pianese, Matthias Nie{\ss}ner, and Luisa Verdoliva.
\newblock {Audio-Visual Person-of-Interest DeepFake Detection}.
\newblock In {\em IEEE Conference on Computer Vision and Pattern Recognition (CVPR) Workshops}, pages 943--952, 2023.

\bibitem{cozzolino2023raising}
Davide Cozzolino, Giovanni Poggi, Riccardo Corvi, Matthias Nie{\ss}ner, and Luisa Verdoliva.
\newblock {Raising the Bar of AI-generated Image Detection with CLIP}.
\newblock {\em IEEE Conference on Computer Vision and Pattern Recognition (CVPR) Workshops}, 2024.

\bibitem{cozzolino2021idreveal}
Davide Cozzolino, Andreas R{\"{o}}ssler, Justus Thies, Matthias Nie{\ss}ner, and Luisa Verdoliva.
\newblock {ID-Reveal: Identity-aware DeepFake Video Detection}.
\newblock In {\em IEEE International Conference on Computer Vision (ICCV)}, pages 15108--15117, 2021.

\bibitem{cuccovillo2023audio}
Luca Cuccovillo, Milica Gerhardt, and Patrick Aichroth.
\newblock {Audio Spectrogram Transformer for Synthetic Speech Detection via Speech Formant Analysis}.
\newblock In {\em IEEE international workshop on information forensics and security (WIFS)}, pages 1--6, 2023.

\bibitem{Van2008visualizing}
Laurens~Van der Maaten and Geoffrey Hinton.
\newblock {Visualizing data using t-SNE}.
\newblock {\em Journal of machine learning research}, 9(11):2579--2605, 2008.

\bibitem{desplanques2020ecapa}
Brecht Desplanques, Jenthe Thienpondt, and Kris Demuynck.
\newblock {ECAPA-TDNN: Emphasized Channel Attention, propagation and aggregation in TDNN based speaker verification}.
\newblock In {\em Annual Conference of the International Speech Communication Association (INTERSPEECH)}, pages 3830--3834, 2020.

\bibitem{Dhamyal2021using}
Hira Dhamyal, Ayesha Ali, Ihsan~Ayyub Qazi, and Agha~Ali Raza.
\newblock {Using Self Attention DNNs to Discover Phonemic Features for Audio Deep Fake Detection}.
\newblock In {\em IEEE Automatic Speech Recognition and Understanding Workshop (ASRU)}, pages 1178--1184, 2021.

\bibitem{ding2023samo}
Siwen Ding, You Zhang, and Zhiyao Duan.
\newblock {SAMO: Speaker Attractor Multi-Center One-Class Learning for Voice Anti-Spoofing}.
\newblock In {\em IEEE International Conference on Acoustics, Speech and Signal Processing (ICASSP)}, pages 1--5, 2023.

\bibitem{guzhov2021esresne}
Andrey Guzhov, Federico Raue, J{\"o}rn Hees, and Andreas Dengel.
\newblock {Esresne (x) t-fbsp: Learning robust time-frequency transformation of audio}.
\newblock In {\em International Joint Conference on Neural Networks (IJCNN)}, pages 1--8, 2021.

\bibitem{guzhov2022audioclip}
Andrey Guzhov, Federico Raue, J{\"o}rn Hees, and Andreas Dengel.
\newblock {AudioCLIP: Extending CLIP to Image, Text and Audio}.
\newblock In {\em IEEE International Conference on Acoustics, Speech and Signal Processing (ICASSP)}, pages 976--980, 2022.

\bibitem{jung2022aasist}
Jee-weon Jung, Hee-Soo Heo, Hemlata Tak, Hye-jin Shim, Joon~Son Chung, Bong-Jin Lee, Ha-Jin Yu, and Nicholas Evans.
\newblock {AASIST: Audio Anti-Spoofing using Integrated Spectro-Temporal Graph Attention Networks}.
\newblock In {\em IEEE International Conference on Acoustics, Speech and Signal Processing (ICASSP)}, pages 6367--6371, 2022.

\bibitem{koluguri2022titanet}
Nithin~Rao Koluguri, Taejin Park, and Boris Ginsburg.
\newblock {TitaNet: Neural Model for speaker representation with 1D Depth-wise separable convolutions and global context}.
\newblock In {\em IEEE International Conference on Acoustics, Speech and Signal Processing (ICASSP)}, pages 8102--8106, 2022.

\bibitem{muller2022does}
Nicolas M{\"u}ller, Pavel Czempin, Franziska Diekmann, Adam Froghyar, and Konstantin B{\"o}ttinger.
\newblock {Does Audio Deepfake Detection Generalize?}
\newblock In {\em Annual Conference of the International Speech Communication Association (INTERSPEECH)}, pages 2783--2787, 2022.

\bibitem{ojha2023towards}
Utkarsh Ojha, Yuheng Li, and Yong~Jae Lee.
\newblock Towards universal fake image detectors that generalize across generative models.
\newblock In {\em IEEE Conference on Computer Vision and Pattern Recognition (CVPR)}, pages 24480--24489, 2023.

\bibitem{Oneata2023towards}
Dan Oneata, Adriana Stan, Octavian Pascu, Elisabeta Oneata, and Horia Cucu.
\newblock Towards generalisable and calibrated synthetic speech detection with self-supervised representations.
\newblock {\em arXiv preprint arXiv:2309.05384v1}, 2023.

\bibitem{pianese2022deepfake}
Alessandro Pianese, Davide Cozzolino, Giovanni Poggi, and Luisa Verdoliva.
\newblock Deepfake audio detection by speaker verification.
\newblock In {\em IEEE international workshop on information forensics and security (WIFS)}, pages 1--6, 2022.

\bibitem{radford2021learning}
Alec Radford, Jong~Wook Kim, Chris Hallacy, Aditya Ramesh, Gabriel Goh, Sandhini Agarwal, Girish Sastry, Amanda Askell, Pamela Mishkin, Jack Clark, et~al.
\newblock Learning transferable visual models from natural language supervision.
\newblock In {\em International Conference on Machine Learning (ICML)}, pages 8748--8763, 2021.

\bibitem{Salvi2023towards}
Davide Salvi, Paolo Bestagini, and Stefano Tubaro.
\newblock {Towards Frequency Band Explainability in Synthetic Speech Detection}.
\newblock In {\em IEEE European Signal Processing Conference (EUSIPCO)}, pages 620--624, 2023.

\bibitem{Shim2022baseline}
Hye-Jin Shim, Hemlata Tak, Xuechen Liu, Hee-Soo Heo, Jee-Weon Jung, Joon~Son Chung, Soo-Whan Chung, Ha-Jin Yu, Bong-Jin Lee, Massimiliano Todisco, et~al.
\newblock {Baseline Systems for the First Spoofing-Aware Speaker Verification Challenge: Score and Embedding Fusion}.
\newblock In {\em Odyssey - The Speaker and Language Recognition Workshop}, 2022.

\bibitem{Tak2021end2end-spectro}
Hemlata Tak, Jee-weon Jung, Jose Patino, Madhu Kamble, Massimiliano Todisco, and Nicholas Evans.
\newblock {End-to-End Spectro-Temporal Graph Attention Networks for Speaker Verification Anti-Spoofing and Speech Deepfake Detection}.
\newblock In {\em Automatic Speaker Verification and Spoofing Countermeasures Challenge}, 2021.

\bibitem{Tak2021end}
Hemlata Tak, Jose Patino, Massimiliano Todisco, Andreas Nautsch, Nicholas Evans, and Anthony Larcher.
\newblock {End-to-end anti-spoofing with RawNet2}.
\newblock In {\em IEEE International Conference on Acoustics, Speech and Signal Processing (ICASSP)}, pages 6369--6373, 2021.

\bibitem{wang2020asvspoof}
Xin Wang, Junichi Yamagishi, Massimiliano Todisco, H{\'e}ctor Delgado, Andreas Nautsch, Nicholas Evans, Md~Sahidullah, Ville Vestman, Tomi Kinnunen, Kong~Aik Lee, et~al.
\newblock {ASVspoof 2019: A large-scale public database of synthesized, converted and replayed speech}.
\newblock {\em Computer Speech \& Language}, 64:101114, 2020.

\bibitem{laionclap2023}
Yusong Wu, Ke~Chen, Tianyu Zhang, Yuchen Hui, Taylor Berg-Kirkpatrick, and Shlomo Dubnov.
\newblock {Large-scale Contrastive Language-Audio Pretraining with Feature Fusion and Keyword-to-Caption Augmentation}.
\newblock In {\em IEEE International Conference on Acoustics, Speech and Signal Processing (ICASSP)}, pages 1--5, 2023.

\bibitem{yamagishi2021asvspoof}
Junichi Yamagishi, Xin Wang, Massimiliano Todisco, Md~Sahidullah, Jose Patino, Andreas Nautsch, Xuechen Liu, Kong~Aik Lee, Tomi Kinnunen, Nicholas Evans, et~al.
\newblock {ASVspoof 2021: accelerating progress in spoofed and deepfake speech detection}.
\newblock In {\em Automatic Speaker Verification and Spoofing Coutermeasures Challenge}, 2021.

\bibitem{Zhang2021one-class}
You Zhang, Fei Jiang, and Zhiyao Duan.
\newblock {One-class Learning Towards Synthetic Voice Spoofing Detection}.
\newblock {\em IEEE Signal Processing Letters}, 28:937--941, 2021.

\end{thebibliography}
\end{document}